\documentclass{iaus}
\usepackage{graphicx}

\title[Herschel observations of PNe] {Herschel observations of PNe\\ in the
  MESS key program
}

\author[P.A.M. van Hoof et al.]
{P.A.M.~van~Hoof$^1$, M.J.~Barlow$^2$,
G.C.~Van~de~Steene$^1$, K.M.~Exter$^3$,
R.~Wesson$^2$, R.~Ottensamer$^4$,
T.L.~Lim$^5$, B.~Sibthorpe$^6$,
M.~Matsuura$^2$, T.~Ueta$^7$, H.~Van~Winckel$^3$, C.~Waelkens$^3$ \\ \and the MESS consortium}

\affiliation{$^1$Royal Observatory of Belgium, Ringlaan 3, B-1180 Brussels, Belgium\\[\affilskip]
$^2$Dept.\ of Physics \& Astronomy, Univ.\ College London, Gower St, London WC1E 6BT, UK\\[\affilskip]
$^3$IvS, Katholieke Universiteit Leuven, Ce\-les\-tij\-nenlaan 200 D, B-3001 Leuven, Belgium\\[\affilskip]
$^4$University Vienna, Dept.\ of Astronomy, T\"urkenschanzstrasse 17, A-1180 Wien, Austria\\[\affilskip]
$^5$Space Science and Technology Dept., Rutherford Appleton Lab., Oxfordshire, OX11 0QX, UK\\[\affilskip]
$^6$UKATC, Royal Observatory Edinburgh, Blackford Hill, Edinburgh EH9 3HJ, UK\\[\affilskip]
$^7$Dept.\ of Physics and Astronomy, Univ.\ of Denver, Mail Stop 6900, Denver, CO 80208, USA}

\pubyear{2012}
\volume{283}
\pagerange{1-4}
\jname{Planetary Nebulae: an Eye to the Future}
\editors{A.C. Editor, B.D. Editor \& C.E. Editor, eds.}

\def\arcsec{$^{\prime\prime}$}

\begin{document}

\maketitle

\begin{abstract}
In this paper we give a progress report on the Herschel observations of
planetary nebulae that are carried out as part of the MESS guaranteed time key
program.
\keywords{planetary nebulae: individual (NGC 6720, NGC 650, NGC 6853, NGC
  7027) --- dust, extinction --- molecules --- infrared: ISM}
\end{abstract}

\firstsection 

\section{Introduction}

As part of the MESS (Mass loss of Evolved StarS) Herschel guaranteed time key
program (PI Martin Groenewegen) we are observing a sample of planetary nebulae
(PNe) in photometric and/or spectroscopic mode with the PACS (\cite[Poglitsch
  et al.\ 2010]{Po10}) and SPIRE (\cite[Griffin et al.\ 2010]{Gr10})
instruments on board the Herschel satellite (\cite[Pilbratt et al.
  2010]{Pi10}). The aims of the MESS program are threefold. 1) Study the time
dependence of the mass loss process via a search for shells and multiple
shells. 2) Study the dust and gas chemistry as a function of progenitor mass.
3) Study the properties and asymmetries of a representative sample of evolved
objects. The program covers many phases of stellar evolution: AGB \& post-AGB
stars, PNe, massive stars, and supernovae. It is described in
\cite[Groenewegen et al.\ (2011)]{Groen11}. In Table~\ref{tab1} we present an
overview of the Herschel observations that have been taken or are planned for
the PNe in our sample.

All targets have been imaged in scan map (SM) mode, except NGC 7293 which was
observed in parallel mode (PM). With PACS we have obtained images in the 70
and 160 $\mu$m bands (beam size: 5.2\arcsec\ and 12\arcsec, respectively),
with SPIRE we have obtained all 3 bands: 250, 350, and 500 $\mu$m (beam size:
18.1\arcsec, 25.2\arcsec, and 36.6\arcsec, respectively). For all
spectroscopic targets we obtain full spectral scans, covering the wavelength
range 51--98 + 103--192 $\mu$m for PACS and 196--667 $\mu$m for SPIRE.

\section{Results}

Below we will discuss preliminary results of selected objects of our sample.

\begin{figure}
\begin{center}
\includegraphics[width=\textwidth]{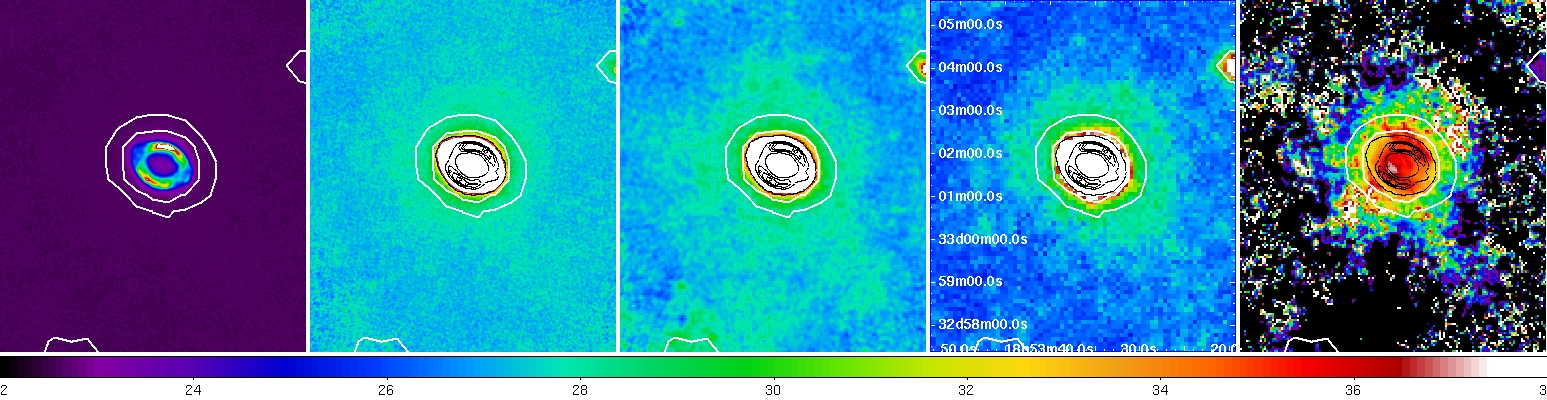}
 \caption{NGC~6720, from left to right: PACS~70~$\mu$m inner and outer region,
   PACS~160$\mu$m, SPIRE~250$\mu$m, and a temperature map created from the
   PACS 70 / 160 $\mu$m ratio image. The black contours are of the
   PACS~70~$\mu$m inner region and the white contours of the fainter outer
   regions of the PACS~160~$\mu$m image. The bar the the bottom shows the
   temperature scale.}
   \label{fig6720}
\end{center}
\end{figure}

{\underline{\it NGC 6720 (the Ring nebula)}}. This PN (shown in
Fig.~\ref{fig6720}) is very similar to NGC 7293 (the Helix nebula). It seems
they follow the same evolutionary path. The Helix nebula has very strong H$_2$
emission \cite[(Storey et al. 1987)]{st87}. A static photoionization model
cannot explain this emission, but a hydrodynamic model can (\cite[Henney et
  al.\ 2007]{Heney07}). This model indicates that the erosion of the knots by
the radiation field of the central star is substantial: between 10$^{-10}$ and
10$^{-9}$ M$_\odot$\,yr$^{-1}$ despite the low luminosity of the central star
(120~L$_\odot$). Considering the fact that the central star luminosity was
much higher in the past and the knots must have been closer to the central
star, survival of the knots from the AGB phase (as was e.g. proposed by
\cite[Matsuura et al. 2009]{Matsuura09}) to the current time seems
problematic. However, more detailed modeling is warranted.

We have developed a photoionization model of the Ring nebula with Cloudy, last
described by \cite[Ferland et al. (1998)]{Ferland98}, which we used to
investigate possible formation scenarios for H$_2$. We conclude that the most
plausible scenario is that the H$_2$ resides in high density knots which were
formed after the recombination of the gas started when the central star
luminosity dropped steeply around 1000-2000 years ago. The models show that
H$_2$ formation in the knots is expected to be substantial since then, and may
well still be ongoing at this moment. For a more detailed discussion see
\cite[van Hoof et al. (2010)]{vHoof10}.

\begin{table}
\begin{center}
\caption{PN observations in the MESS program. SM stands for scan-map imaging,
PM stands for parallel mode imaging, and RS stands for full range spectroscopy.
Observations marked ``tbd'' still need to be carried out, for others we give
the observing date in ISO format.}
\label{tab1}
{\scriptsize
\begin{tabular}{ll}
\hline
target & observation type (date) \\
\hline
NGC 650 & PACS SM (2010-02-23), SPIRE SM (2010-08-24) \\
NGC 3587 & PACS SM (2010-06-21), SPIRE SM (2010-05-08) \\
NGC 6302 & PACS RS (tbd), SPIRE RS (2010-02-26) \\
NGC 6537 & PACS RS (tbd), SPIRE RS (tbd) \\
NGC 6543 & PACS SM (2011-05-07), SPIRE SM (2009-12-26), PACS RS (2009-06-23) \\
NGC 6720 & PACS SM (2009-10-10), SPIRE SM (2009-10-06) \\
NGC 6853 & PACS SM (2010-04-04), SPIRE SM (2010-10-11) \\
NGC 7027 & PACS SM (2010-05-05), PACS RS (2009-11-12), SPIRE RS (2010-01-09) \\
NGC 7293 & PACS+SPIRE PM (2010-04-28) \\
IRAS 22036+5306 & PACS SM (2010-06-22) \\
\hline
\end{tabular}
}
\end{center}
\end{table}

\begin{figure}
\begin{center}
\includegraphics[width=\textwidth]{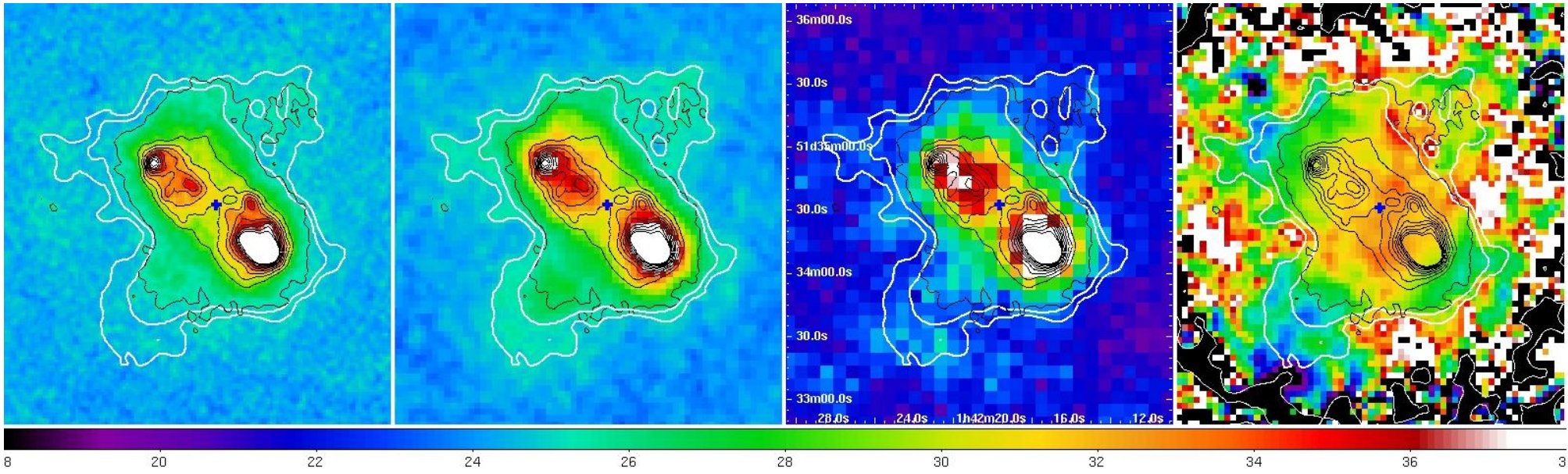} 
 \caption{NGC 650, from left to right: PACS 70 and 160 $\mu$m, SPIRE 250
   $\mu$m, and a temperature map. The blue cross marks the central star. The
   contours were generated the same way as in Fig.~\ref{fig6720}. The bar on
   the bottom shows the temperature scale.}
   \label{fig650}
\end{center}
\end{figure}

{\underline{\it NGC 650 (the Little Dumbbell nebula)}}. This is a bipolar
nebula. We can clearly see the edge-on equatorial density enhancement. The
blobs towards the SE and NW are detected, though faint. The bipolar lobes
themselves are not detected. In Fig.~\ref{fig650} we show the PACS 70 and 160
$\mu$m and SPIRE 250 $\mu$m images, as well as the temperature map of the dust
created from the PACS 70~/ 160 $\mu$m ratio image, after convolution to the
same point spread function (PSF). The black contours are taken from the PACS
70 $\mu$m image to indicate the position of the torus. The shadowing effect of
the torus is clearly visible: the regions at the outer edge of the torus and
beyond are clearly cooler than the dust in other directions. The dust grains
are primarily heated by UV photons, either emitted by the central star, or
diffuse emission from the gas (e.g., Ly$\alpha$ photons). Hence there must be
substantial extinction of UV photons inside the torus, despite the rather low
density of these regions (140--400 cm$^{-3}$, \cite[Minkowski \& Osterbrock
  1960]{MO60}).

\begin{figure}
\begin{center}
\includegraphics[width=\textwidth]{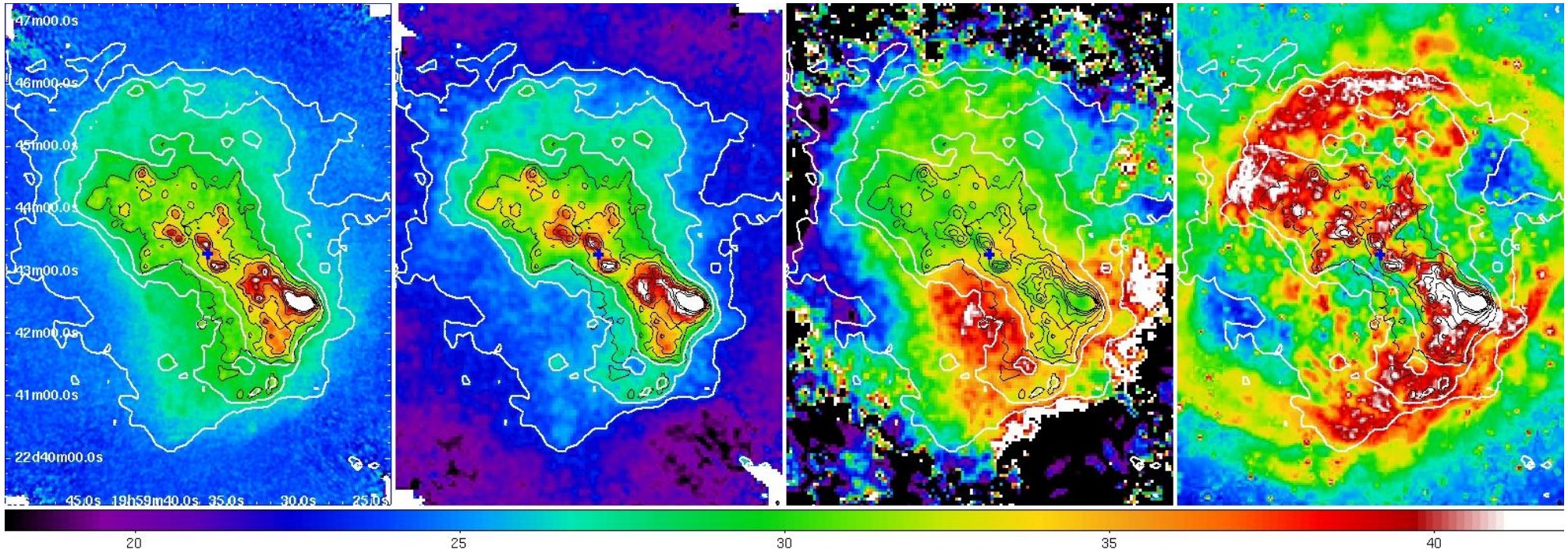} 
 \caption{NGC 6853, from left to right: PACS 70 and 160 $\mu$m, the
   temperature map, and an H$\alpha$ image (Robert Gendler, APOD). The blue
   cross marks the central star. The contours were generated the same way as
   in Fig.~\ref{fig6720}. The bar on the bottom shows the temperature scale.}
   \label{fig6853}
\end{center}
\end{figure}

{\underline{\it NGC 6853 (the Dumbbell nebula)}}. Also in this PN we see a
detailed match between the dust emission in the PACS image and the H$_2$
emission from ground based imaging, showing the close association of the dust
and the H$_2$. In Fig.~\ref{fig6853} we show the PACS images of this source,
as well as the temperature map derived from the PACS 70~/ 160 $\mu$m ratio
image after convolution to the same PSF. The black contours are taken from the
PACS 70 $\mu$m image to indicate the position of the high density regions. It
is clear that there is a strong correlation between the high-density regions
and the colder dust. The hot patch towards the south appears to be real and
has no counterpart in the north. Presumably this is material that is directly
irradiated by the central star.

{\underline{\it NGC 7027}}. In the PACS and SPIRE spectra of this PN we see
atomic lines from H\,{\sc i}, [C\,{\sc i]}, [C\,{\sc ii]}, [N\,{\sc ii]},
[N\,{\sc iii]}, [O\,{\sc i]}, [O\,{\sc iii]} and molecular lines from
$^{12}$CO, $^{13}$CO, OH, H$_2$O, CH, CH$^+$, C$_2$H, HCN, HCO$^+$, and
possibly CN and OH$^+$. Oxygen-rich molecules can be formed because the harsh
radiation field keeps CO partially dissociated throughout the
photo-dissociation region (PDR).

Using the photoionization code Cloudy we have created a preliminary model of
the ionized region and the PDR of NGC 7027. Using either a constant density or
a constant pressure density law, we could not produce a satisfactory fit to
both the ionized region and the PDR. If on the other hand we use a piecewise
constant density law with different densities in the ionized region and the
PDR, we get a much better fit. This model indicates that the density in the
PDR is much higher than in the ionized region (7.9$\times$10$^5$ and
3.3$\times$10$^4$ cm$^{-3}$, respectively). This density contrast is the
result of the heating by the photoionization process, causing the gas to
expand. From our modeling efforts it becomes clear that the gas cannot be in
pressure equilibrium.

\begin{figure}
\begin{center}
\mbox{
\includegraphics[width=0.43\textwidth]{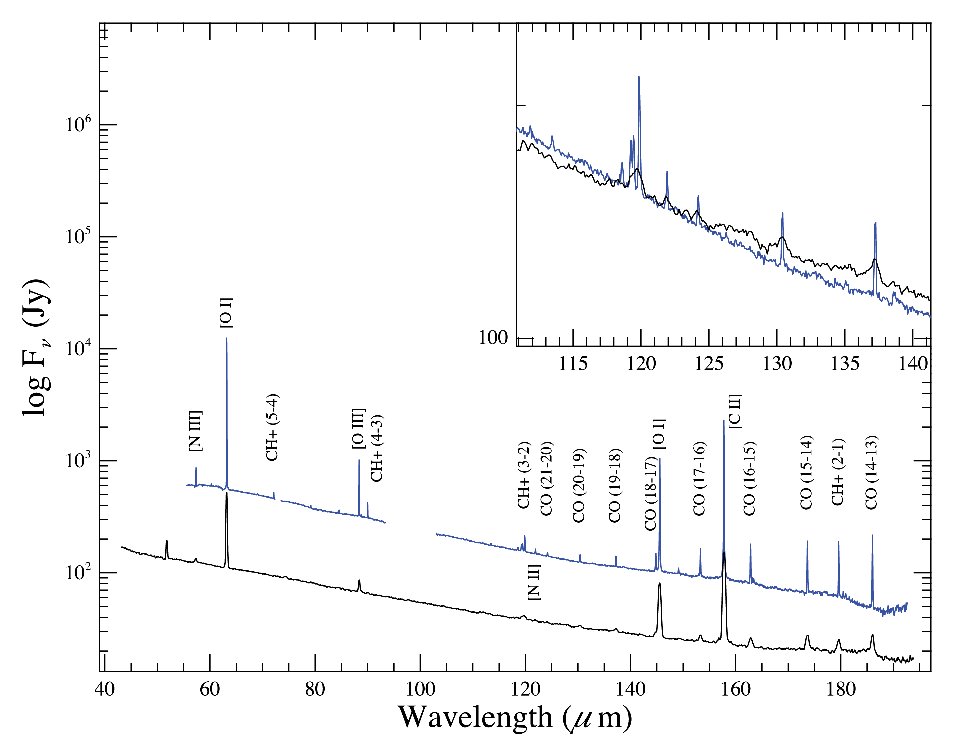}\hspace*{0.07\textwidth}
\includegraphics[width=0.43\textwidth]{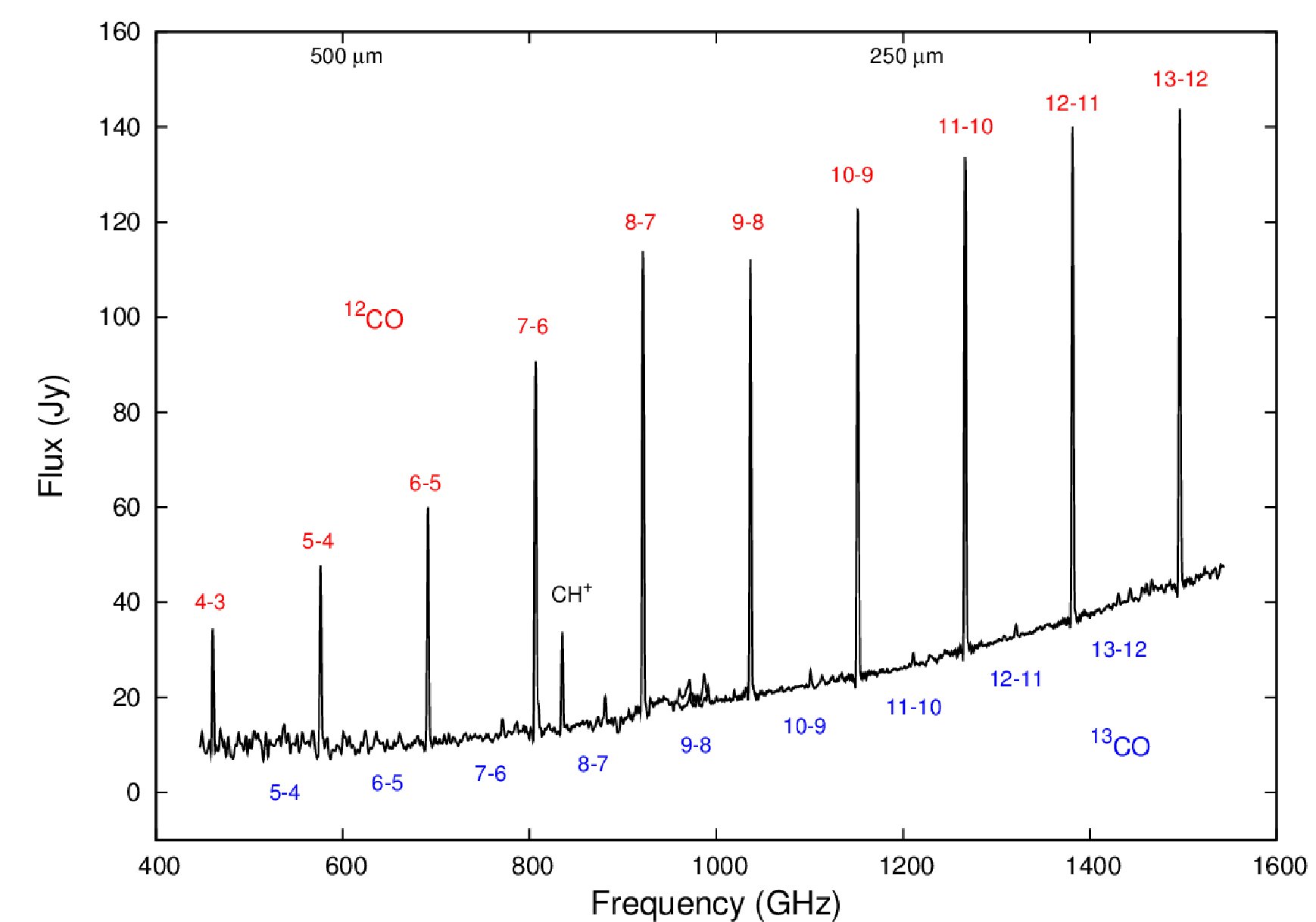} 
}
 \caption{Left: the PACS spectrum, right: the SPIRE spectrum. Underneath the
   PACS spectrum we show the ISO-LWS spectrum for comparison (shifted by an
   arbitrary amount).}
   \label{fig7027}
\end{center}
\end{figure}

\section{Summary}

For most PNe we see a detailed match between the FIR dust emission and the
H$_2$ emission, showing that the dust and H$_2$ are closely associated. In the
case of NGC 6720 this close association provides the first observational
evidence for H$_2$ formation on oxygen-rich dust grains in an astrophysical
environment. Using Cloudy models we conclude that the most likely scenario is
that in NGC 6720 H$_2$ formed (and may still be forming now) inside dense
knots that started to form after recombination of the nebula started 1000-2000
years ago. We produced temperature maps from the PACS 70~/ 160 $\mu$m ratio
images which show a rich structure. They indicate that internal extinction in
the UV is important, despite the highly evolved status of the nebulae. Many
molecules were detected in the spectra of NGC 7027 and NGC 6302. A preliminary
Cloudy photoionization / PDR model of NGC 7027 indicates that the PDR has
higher density than the ionized region and the nebula is not in pressure
equilibrium.

\acknowledgements

{
PvH acknowledges support from Belspo through the ESA PRODEX Programme.
}


\begin{thebibliography}{}

\bibitem[bla]{Ferland98}
Ferland, G.J., Korista, K.T., Verner, D.A. et al., 1998,
\textit{PASP}, 110, 761

\bibitem[bla]{Gr10}
Griffin, M.J., Abergel, A., Abreu, A. et al., 2010,
\textit{A\&A}, 518, L3

\bibitem[bla]{Groen11}
Groenewegen, M.A.T., Waelkens, C., Barlow, M.J. et al. 2011,
\textit{A\&A}, 526, A162

\bibitem[bla]{Henney07}
Henney, W.J., Williams, R.J.R., Ferland, G.J., Shaw, G. \& O'Dell, C.R., 2007,
\textit{ApJ}, 671, L137

\bibitem[bla]{Matsuura09}
Matsuura, M., Speck, A.K., McHunu, B.M. et al., 2009,
\textit{ApJ}, 700, 1067

\bibitem[bla]{MO60}
Minkowski, R. \& Osterbrock, D., 1960,
\textit{ApJ}, 131, 537

\bibitem[bla]{Pi10}
Pilbratt, G.L., Riedinger, J.R., Passvogel, T. et al., 2010,
\textit{A\&A}, 518, L1

\bibitem[bla]{Po10}
Poglitsch, A., Waelkens, C., Geis, N. et al., 2010
\textit{A\&A}, 518, L2

\bibitem[bla]{St87}
Storey, J.W.V., Webster, B.L., Payne, P. \& Dopita, M.A., 1987,
\textit{IAU Symp. 120}, p. 339

\bibitem[bla]{vHoof10}
van Hoof, P.A.M., Van de Steene, G.C., Barlow, M.J. et al., 2010,
\textit{A\&A}, 518, L137

\end{thebibliography}
\end{document}